**A centrosymmetric hexagonal magnet with superstable biskyrmion magnetic nanodomains in a wide temperature range of 100K to 340K**

*Wenhong Wang\*, Ying Zhang\*, Guizhou Xu,  Licong Peng, Bei Ding, Yue Wang, Zhipeng Hou, Xiaoming Zhang, Xiyang Li, Enke Liu, Shouguo Wang, Jianwang Cai, Fangwei Wang, Jianqi Li, Fengxia, Hu, Guangheng Wu, Baogen Shen, Xi-xiang Zhang*

Dr. W. H. Wang, Dr. Y. Zhang, Dr. G. Z. Xu, L. C. Peng, B. Ding, Y. Wang, Z. P. Hou, Dr. X. M. Zhang, X. Y. Li, Dr. E. K. Liu, Prof. J. W. Cai, Prof. F. W. Wang, Prof. J. Q. Li, Prof. F.X. Hu, Prof. G. H. Wu, Prof. B. G. Shen
Beijing National Laboratory for Condensed Matter Physics, Institute of Physics, Chinese Academy of Sciences, Beijing 100190, China
E-mail: wenhong.wang@iphy.ac.cn (W. H. Wang), Zhangy@iphy.ac.cn (Y. Zhang)
Dr. G. Z. Xu
School of Materials Science and Engineering, Nanjing University of Science and Technology, Nanjing 210094, China
Prof. S. G. Wang
School of Materials Science and Engineering, University of Science and Technology Beijing, Beijing 100083, China
Prof. X. X. Zhang
King Abdullah University of Science and Technology (KAUST), Physical Science and Engineering, Thuwal 23955-6900, Saudi Arabia
[*] W.H.W. and Y. Z contributed equally to this work.

Magnetic skyrmions are topologically protected vortex-like nanometric spin textures that have recently received growingly attention for their potential applications in future high-performance spintronic devices. Such unique mangetic naondomains have been recently discovered in bulk chiral magnetic materials, such as MnSi[1-4], FeGe[5,6], FeCoSi[7], $Cu_2OSeO_3$[8-10], β-Mn-type Co-Zn-Mn[11], and also $GaV_4S_8$[12] a polar magnet. The crystal structure of these materials is cubic and lack of centrosymmetry, leading to the existence of Dzyaloshinskii-Moriya (DM) interactions. Unlike the conventional spin configurations, such as helical or conical, that are usually found in chiral magnets, a magnetic skyrmion has a particle-like swirling-spin configuration characterized by a topological index called the skyrmion number[13,14]. The nontrivial topology of magnetic skyrmions results in a number of intriguing electromagnetic phenomena, including the topological Hall effect[2], skyrmion magnetic resonance[15] and thermally induced ratchet motion[16]. These topologically protected properties, together with their nanoscale dimensions (10 ~ 100 nm) and ultralow threshold for current-driven motion ($10^5 \sim 10^6$ Am$^{-2}$) [17-19], make the magnetic skyrmions be promising materials for technological applications in spincronic devices[20].

In addition to the observation of magnetic skyrmions in chiral magnets characterized by DM interactions, magnetic skyrmions have also been recently identified in centrosymmetric magnets with uniaxial magnetic anisotropy[21]. In particular, magnetic biskyrmions (or skyrmion molecules), a new type of skyrmion formed by two skyrmions of opposite spin helicities, have been found in a centrosymmetric bilayered magnetite[22]. Compared with a single skyrmion in cubic chiral-lattice magnets, a unique feature of such biskyrmions is that their topological charges and sizes can be controlled by the thickness of the sample, the external magnetic field and even the electrical current, which render them promising materials for alternative information storage media in spintronics applications. However, experimental observations of biskyrmions have, thus far, been limited to bilayered magnetites at temperatures below 60 K, making their use in room temperature (RT) spintronic devices impracticable. Finding new materials that host magnetic biskyrmions at RT is thus crucial to their further development and use.

One possible path to the discovery of new centrosymmetric magnets that can host skyrmions near RT is to revisit non-cubic, noncollinear magnets that have been reported to be ferromagnetic above RT. As in bilayered magnetite[22], the magnetic dipolar interaction combined with the low crystal symmetry in such magnets might provide additional stability to skyrmions. In what follows, we describe a hexagonal magnet, MnNiGa, that has a layered Ni$_2$In-type structure [23] (space group *P63/mmc*) with alternating stacks of Mn atomic layers and Ni-Ga atomic layers along the *c*-axis (**Figure 1a**). In our previous work [24], we have established a stable magnetic-structural coupling and realized a giant magnetocaloric effects in another $Ni_2In$-type hexagonal MnNiGe:Fe compounds. Moreover, we have fabricated the off-stoichiometry sputtered MnNiGa hexagonal film, which showed some abnormal transport properties[25]. For the $Ni_2In$-type MnNiGa compound, the 2(a) site is preferentially occupied by Mn atoms, while the 2(d) is mainly occupied by Ni atoms, in off-stoichiometric cases, extra Mn atoms can also occupy this site, and the 2(c) site is occupied by Ga atoms. Previous powder neutron diffraction exeepriments[26] revealed that the magnetic structure varies with the Mn concentration, causing the spin-orientation to shift with changing temperature.

Here, we report on the experimental discovery of biskyrmion magnetic nanodomains at RT and the observation of a biskyrmion-derived topological Hall effect (THE) in the centrosymmetric hexagonal MnNiGa magnet. Using a phase reconstruction technique based on a transport-of-intensity equation (TIE), we established the texture of the biskyrmion spin. Results from Lorentz transmission electron microscopy (TEM) and the topological Hall effect revealed that the biskyrmion phase is stable over a much wider temperature range (100 K to ~340K) and a larger magnetic field range in our material than in skyrmion-hosting bulk crystals reported previously[1-3,6]. The super-wide operating temperature and the broad range of material options indicate important progress toward the realization of skyrmion-based spintronic devices.

We synthesized a series of polycryatalline $(Mn_{1-x}Ni_x)_{65}Ga_{35}$ ($0.15 \leq x \leq 0.5$) alloys by induction melting pure metals in an argon atmosphere (see methods). Powder X-ray diffraction measurements and Rietveld refinements indicated that all as-prepared samples crystallized into a $Ni_2In$-type structure. Upon substituting Ni with Mn, the *c*-axis increased and the *a*-axis decreased (**Figure S1, SI**). The Curie temperature, $T_c$, systematically decreased with increasing x with a minimum at x = 0.2, which is consistent with the results reported by Shiraishi *et al.*[26]. Uniform magentic stripe structures are observed in most $(Mn_{1-x}Ni_x)_{65}Ga_{35}$ ($0.15 \leq x \leq 0.5$) alloys (**Figure S2, SI**). In the present study, we focus only on the (Mn$_{1-}$ $_x$Ni$_x$)$_{65}$Ga$_{35}$ alloy with x = 0.5, since it has the highest $T_c$ at ~ 350 K with a shorter stripe periodicity (**Figure 1c,d**). According to previous powder neutron diffraction measurements in off-stoichiometric MnNiGa (ref. 26), the Mn spin arrangement at the 2(a) site is a canted one, with ferromagnetic components parallel to the *c*-axis. Thus, the canting of the two Mn spins towards the *c*-axis is considered to be the origin of the weak magnetic anisotropy in this material system. Moreover, as shown in **Figure 1c**, a hump is observed at temperature $T^*$ ~200 K, which indicates the appearance of a new magnetic structure but remain a topic for future investigations.

**Figure 2** shows under-focused Lorentz TEM images of revolving skyrmions with different magnetic fields in (Mn$_{1-x}$Ni$_x$)$_{65}$Ga$_{35}$ alloy with x=0.5 measured at $T$=300 K. At $\mu_0 H$=0 T (**Figure 2a**), a spontaneous magnetic stripe structure with a periodicity of $\lambda$~180 nm is observed. In **Figure 2b**, we present the corresponding spin texture obtained by using the TIE. The green and red regions shown in **Figure 2b** correspond to in-plane anti-parallel magnetic domains, while the black regions represent the out-of plane magnetic domains. The revolving behavior of the biskyrmions under a vertical magnetic field is similar to that reported previously[5,7,8,21,22]. The dynamics of formation and disappearance of the biskyrmions with increasing applied magnetic field were successfully recorded by using *in-situ* Lorentz TEM (**see Supplementary Movie, SI**). The threshold magnetic field of skyrmion formation depends on the sample thickness and the temperature. With increasing magnetic field, the magnetic stripes narrow first and then break into fragments at the locations where the stripes narrow faster (**Figure 2 c-f**). Simultaneously, the biskyrmions domains begin to appear. As the magnetic field increases further to 0.28T (**Figure 2g,h**), the magnetic stripes are completely replaced by individual biskyrmions domains. When the magnetic field is stronger than 0.3T, the size of the biskyrmions domains begins to shrink and they eventually vanish at a critical magnetic field of 0.4T. This critical magnetic field is likely a function of the temperature and sample thickness, closely correlated to the saturation magnetic field at which a uniform ferromagnetic domain is formed.

**Figure 3** shows part of magnified images of **Figure 2e**. The opposite magnetic contrast between the under-focused (**Figure 3a**) and over-focused (**Figure 3b**) Lorentz TEM images of the biskyrmions is clearly different from the contrast observed in traditional skyrmions with $N_s$=1. **Figure 3c** shows the corresponding spin textures of the highlighted area in **Figure 2f** obtained by TIE analyses of the Lorentz TEM data. To reveal the spin texture of biskyrmions, we show magnified, over-focused and under-focused Lorentz TEM images of a biskyrmions in **Figure 3e and 3f**, respectively. The corresponding TIE analysis (**Figure 3d**) clearly indicates that each spin configuration is composed of two single skyrmions with opposite magnetic helicities; that is, each skyrmion has clockwise (CW) and counter-clockwise (CCW) spin curls with an out-of-plane magnetization culminating at the core. This spin configuration is the same as that of biskyrmions reported by Yu et al.[22] with $N_s$=2. We define the polarity of constituent skyrmions as plus (+) for CW skyrmions and minus (-) for CCW skyrmions. The formation of biskyrmions stems from the helicity degree of freedom (+ or -) of each skyrmion, which can be realized only in centrosymmetric magnets and not in chiral-lattice magnets with DM interactions. However, in contrast with the random polarity of skyrmions observed in another centrosymmetric magnet [21], the polar direction of the biskyrmion described here is fixed in MnNiGa, which agrees well with previous observations in bilayered magnetites [22].

In addition to direct observations from Lorentz TEM, magnetic transport measurements, especially observations of the topological Hall effect (THE), have been considered as a hallmark for the existence of the spin chirality, e.g. the skyrmion spin textures in helimagnets with B20 cyrstal structure [2, 27-30]. In **Figure 4a**, we show the Hall resistivities, $\rho_{xy}$, measured at various temperatures from 340 K to 10 K. In the presence of a topological spin texture, the total Hall resistivity can usually be expressed as the sum of various contributions: $\rho_{xy} = R_0 H + R_S M + \rho_{xy}^T$, where $R_0$ is the ordinary Hall coefficient, $R_S$ is the anomalous Hall coefficient, $\rho_{xy}^T$ is the topological Hall resistivity as possible induced by the biskyrmion spin textures, $H$ is magnetic field perpendicular to the sample plane, and $M$ is the corresponding magnetization. To extract THE quantitatively, we singled out both the ordinary and anomalous Hall resistivities from $\rho_{xy}$. As shown in **Figure 4c**, the perpendicular magnetoresistance is very small (less than 3%) and $R_S$ is nearly constant with a magnetic field. Therefore, as in previous studies [29], $R_S$ can simply be determined from $R_S = S_A \rho_{xx}^2$, where $S_A$ is independent of the magnetic field. Here, we define the kink field as the critical field $H_c$. At high fields ($H > H_c$), $\rho_{xy}^T$ is supposed to be zero because all spins are aligned. Hence, combining with field-dependent magnetization M(H) curves measured at different temperatures (**Figure S3, SI**), we determine the $R_0$ and $S_A$ from the intercept and the slope of the linear fitting of $\rho_{xy}/H$ vs. $\rho_{xx}^2 M/H$ in high magnetic field regions (**Figure S4, SI**). We can thus extract $\rho_{xy}^T$ from the difference between the total Hall resistivity $\rho_{xy}$, and the fitted curve $R_0 H + R_S M$ at $H < H_c$. In the inset of **Figure 4a**, we show the representative $\rho_{xy}$-$H$ curve (black line) measured at 250 K, calculated $R_0 H + R_S M$ curve (red line) and the derived $\rho_{xy}^T$ curve (blue line) from the analysis. At $H > H_c$, the calculated result is in accord with the experimental data (**Figure S5, SI**). Strikingly, a hump-like maximum in the curves of $\rho_{xy}^T$ as a function of the applied magnetic field is evident, which was assumed to be a unique signature of THE [2,28,30]. The $\rho_{xy}^T$ values obtained from the experimental data measured over the temperature range of 10 K to 350 K are shown in **Figure 4b**, where a sizable $\rho_{xy}^T$ is clearly found over the whole temperature range. We note that $R_0$ is positive at all temperatures, indicating hole-like conduction in MnNiGa. However, the sign of $R_s$ is positive at high temperatures and changes to negative at temperatures below 100 K.

Furthermore, the perpendicular magnetoresistance data measured over a broad temperature range (10 K to 340 K) in the MnNiGa alloy shows three apparent kinks (highlighted by dashed lines in **Figure 4c**), indicative of three critical magnetic fields, identified as $H_a$, the transition field from the helical to the biskyrmion state, $H_m$, the magnetic field of the pure biskyrrmion state, and $H_c$, the transition field from the biskyrmion to the full ferromagnetic state. These critical magnetic fields were previously identified in magnetoresistance curves of bulk MnSi [32] and nanowire samples [19,33].

Based on the temperature-dependent Lorentz TEM images (**Figure S6 and Table S1, SI**), magnetoresistance data and Hall effect results obtained from MnNiGa thin plates, we created biskyrmion phase diagrams (**Figure 4d**) as represented by contour mapping of the derived $\rho_{xy}^T$ value in the temperature ($T$) and magnetic field ($H$) planes. The biskyrmion phase indicated by the emergence of the THE extends across a broad region in the $T$-$H$ plane, spanning from 340K to nearly 50K and occupying almost all the region below $H_c$ except the near-zero magnetic field. Within such a broad temperature range, as shown in inset of **Figure 4d**, the sign of $\rho_{xy}^T$ is positive and varies systematically and reaches a maximum value of $\rho_{xy}^T \approx$ 0.15μΩcm in the vicinity of 200K. This result is consistent with the temperature dependence we observed in the Lorentz TEM images, as shown in **Figure S6**, in which the biskyrmion density is roughly at a maximum at $T$=215K. This maximum value of the observed $\rho_{xy}^T$ (~0.15μΩcm) for MnNiGa is almost same as that for bulk MnGe (~0.16μΩ cm)[31], which has a largest $\rho_{xy}^T$ value among the bulk B20-type chiral magnets. The THE resistivity resulting from the fictitious magnetic field generated by the skyrmion spin textures can be approximated with $\rho_{xy}^T \approx PR_0B_{eff}^z$ [2], where $B_{eff}^z$ is the fictitious effective field and $P$ is the local spin polarization of the charge carriers. Our band structure calculations indicated that the net spin polarization at the Fermi surface is around 0.33 for $(Mn_{1-x}Ni_x)_{65}Ga_{35}$ with x=0.5 (**Figure S7, SI**). For MnNiGa, the estimated $B_{eff}^z$ is about 1.1 T if we take the biskyrmion size as 90 nm with topological number of 2[29]. Taking into account the value of $R_0 = 0.036 \mu\Omega cm/T$ measured at T=340 K for MnNiGa, we obtained a value of $\rho_{xy}^T \approx 0.012 \mu\Omega cm$. This value is in good agreement with the experimental value of $0.01 \mu\Omega cm$. But with lowering temperature, the value of THE dramatically increases and reaches a maximum value of $0.15 \mu\Omega cm$ at T=200 K. This inconsistency may be originated from the new spin texture appears at temperatures around 200 K as we mentioned above, which can cause potential error for extracting THE signal.

The formation of biskyrmion magnetic nanodomains in a hexagonal $(Mn_{1-x}Ni_x)_{65}Ga_{35}$ with x=0.5 magnet was clearly demonstrated by direct real-space Lorentz TEM observations and confirmed indirectly by THE. In striking contrast to other skyrmion-hosting crystals in which magneic skyrmion nanodomians are stable only in a narrow temperature range, as shown in **Figure 5**, the biskyrmion states in MnNiGa are stabilized over an extremely wide temperature and magnetic field range. In addition, our discovery suggests that a potentially huge family of non-cubic, centrosymmertic crystals will lead to deeper exploration of novel skyrmion-hosting materials. The very broad temperature range and magnetic field range in which the biskyrmions stably existed are extremely important for both fundamental research and potential applications in novel spintronic devices. The stability of skyrmion lattices over extended areas in their ground state at room temperature has been observed in patterned ultrathin single-layer and multilayer magnetic films with perpendicular magentic anisotropy[34-38]. In these systems, both the "intrinsic" interfacial DM interactions and the defect-related "extrinsic" factors (e.g., edge defects and surface roughness) played important roles in inducing magnetic vortex states. More theoretical studies are needed to determine if the biskyrmion phase originated from the magnetic dipolar interaction or other unknown mechanisms related to the defects. To address this issue, further small-angle neutron scattering (SANS) experiments are needed to shed light on the bulk nature of the magnetic structures of MnNiGa with vary compositions. Since such a topological spin structure can be realized in polycrystalline materials in an extremely wide temperature range, our findings not only open up a new path way for further experimental and theoretical studies of quantum magneto-transport properties, e.g., the quantum Hall effect and topological Hall effect, but they may also lead to the realization of skyrmion-based spintronic devices.

**Experimental Section**

*Sample preparation*: A series of polycrystalline $(Mn_{1-x}Ni_x)_{65}Ga_{35}(0.15 \leq x \leq 0.5)$ alloys were synthesized by arc-melting mixtures of highly pure Mn, Ni and Ga metals in a pure argon atmosphere. Excess Mn (2 mol%) over the stoichiometric composition was added to compensate for loss during arc-melting. The crystal structure of the alloys was studied using powder X-ray diffraction measurements and lattice constants were calculated using Rietveld refinements. The composition was analyzed by both energy dispersive X-ray analysis installed in a scanning electron microscopy and inductively coupled plasma spectroscopy.

*Magnetic and Transport measurements*: To measure the (magneto-) transport properties, several polycrystalline crystals were milled into a bar-shape with a typical size of about $3.0 \times 1.0 \times 0.05 mm^3$. Both longitudinal and Hall resistivities were measured using a standard four-probe method on a Quantum Design Physcial Property Measurement System (PPMS). The field dependence of the Hall resistivity was obtained after subtracting the longitudinal resistivity component. The zero-field remnant magnetization, *M,* was also measured using the same field-cooling procedures as used in both longitudinal and Hall resistivity measurements using the Quantum Design PPMS.

***Lorentz TEM measurements***: The thin plates for Lorentz TEM observations were cut from bulk polycrystalline samples and thinned by mechanical polishing and argon-ion milling. The magnetic domain contrast was observed by using Tecnai F20 in the Lorentz TEM mode and a JEOL-dedicated Lorentz TEM, both equipped with liquid-nitrogen, low-temperature holders (~100 K) to study the temperature dependence of the magnetic textures. The magnetic structures could be examined directly in the electron microscope by making use of the Lorentz force experienced by a beam of electrons traversing amagnetic field. The Lorentz TEM is very powerful for observing magnetic configurations in ferromagnetic materials owing to its high spatial resolution. To determine the spin helicity of the skyrmions, the three sets of images with under-, over- and just (or zero) focal-lengths were recorded by a charge coupled device (CCD) camera and then the high-resolution in-plane magnetization distribution map was obtained by the QPt software based on the transport of the intensity equation (TIE). The inversion of magnetic contrast can be seen between the under- and over-focused images. The colors and arrows depict the magnitude and orientation of the in-plane magnetization. The objective lens was turned off when the sample holder was insterted and the perpendicular magnetic field was induced to the stripe domain by increasing the objective lens in small increments. The specimens for the TEM observations were prepared by polishing, dimpling, and subsequently ion milling. The grain size was large enough to be treated as a single grain in TEM. The crystalline orientation for the grain was checked by selected-area electron diffraction (SAED).

***Band structure calculations.***The bulk spin polarization of $(Mn_{1-x}Ni_x)_{65}Ga_{35}$ with x=0.5 were calculated by means of the CASETP code based on a pseudopotential method with a plane-wave basis set [39].The exchange and correlation effects were treated using the generalized-gradient approximation (GGA) [40]. For all cases, an energy cut-off of 400eV for the plane-wave expansion and a Monkhorst-Pack special k-point mesh of $12\times12\times12$ were used for Brillouin zone integrations. The convergence tolerance for the calculations was selected as a difference in total energy within $1 \times 10^{-6}$ eV/atom.

**Supporting Information**

Supporting Information is available from the Wiley Online Library or from the author.

**Acknowledgements**

This work is supported by the National Basic Research Program of China (973 Program 2012CB619405, 2014CB921002), National Natural Science Foundation of China (Grant Nos. 11474343, 51431009, 51590880 and 51471183), and Strategic Priority Research Program B of the Chinese Academy of Sciences under the grant No. XDB07010300.

Received: ((will be filled in by the editorial staff))
Revised: ((will be filled in by the editorial staff))
Published online: ((will be filled in by the editorial staff))

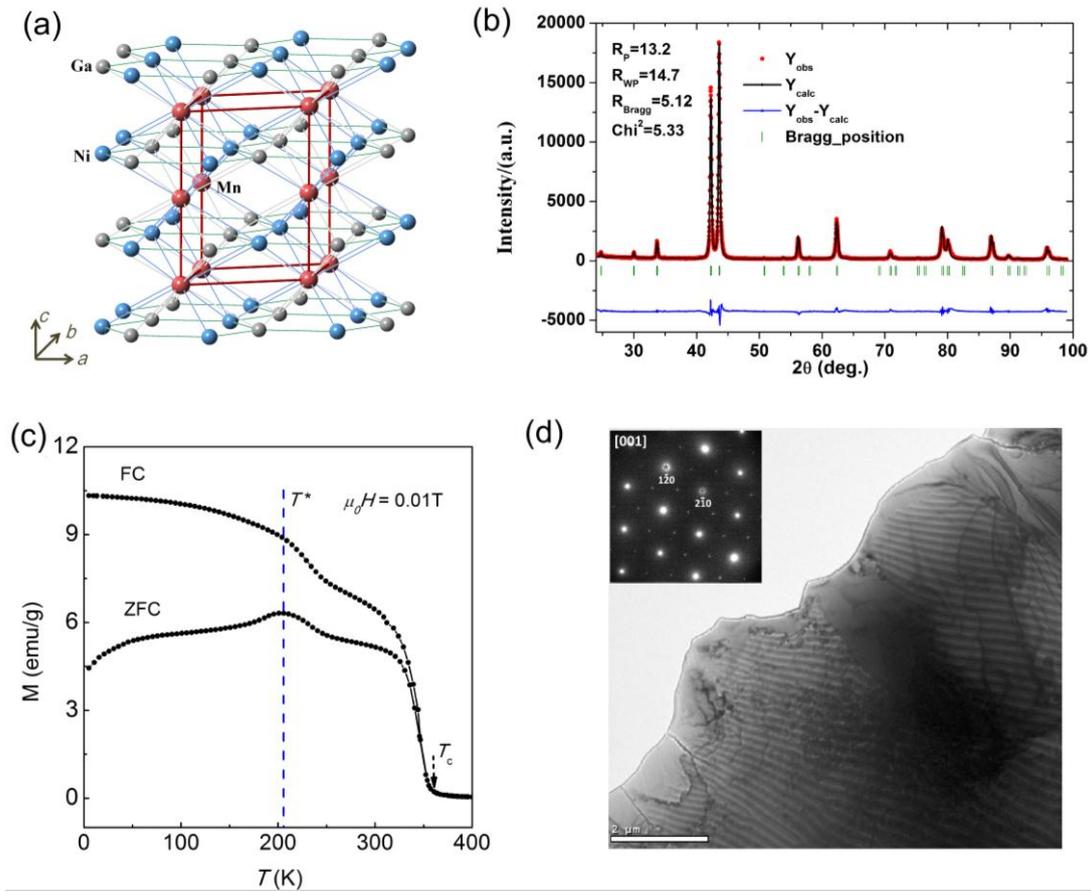

**Figure 1. Structural and magnetic properties of (Mn$_{1-x}$Ni$_x$)$_{65}$Ga$_{35}$ with x = 0.5.** (a) Schematic of a layered Ni$_2$In-type structure Mn-Ni-Ga (space group *P63/mmc*) with alternating stacks of Mn atomic layers and Ni-Ga atomic layers along the *c*-axis. (b) The Rietveld refinement of the powder X-ray diffraction pattern of the sample. (c) Temperature dependence of the zero-field cooling (ZFC) and field-cooling (FC) magnetization curves measured at *μ₀H*=0.01 T. (d) The Lorentz TEM image taken at room temperature shows the uniform magnetic stripe. The scale bar is 2 μm. The inset shows the selected area electron diffraction (SAED), which indicates the [001] orientation.

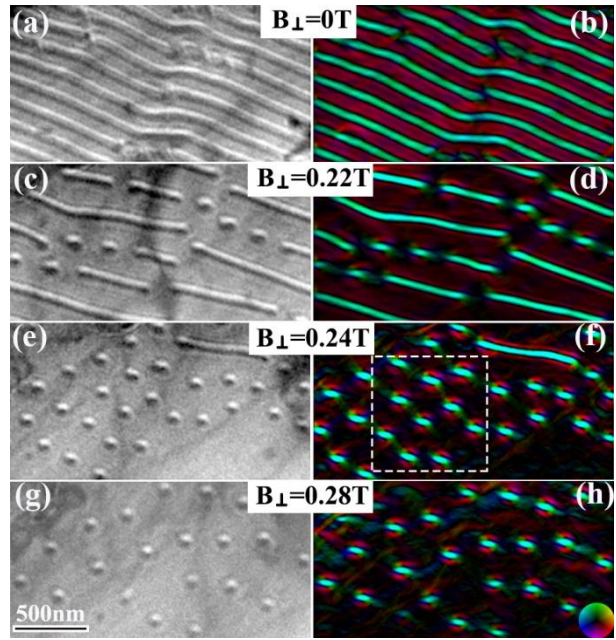

**Figure 2. Magnetic field dependence of biskyrmions in real-space Lorentz TEM images at room temperature.** (a) The under-focused Lorentz TEM image of in-plane spontaneous stripes taken at zero field and the corresponding spin texture (b) obtained by TIE analysis of the Lorentz TEM data. (c) Under-focused Lorentz TEM images of broken stripes and biskyrmions at a magnetic field of 0.22T applied normal to the plane. (e) Lorentz TEM images (under-focused) of dominant biskyrmions at a larger magnetic field of 0.24T. (g) Under-focused Lorentz TEM images of a completely biskyrmion state when the perpendicular magnetic field is increased to 0.28T. (d), (f), and (h) corresponding in-plane magnetization distribution maps obtained by the QPt software based on the transport of intensity equation (TIE). The colors and arrows depict the magnitude and orientation of in-plane magnetizations. The color wheel is in the right corner. The scale bar is 500 nm.

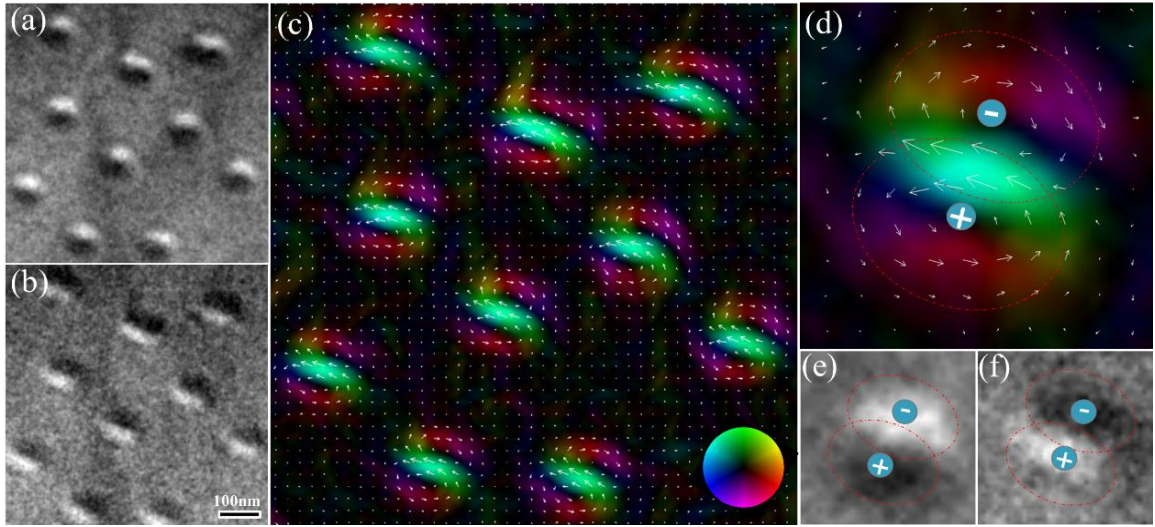

**Figure 3. Magnetic configurations of the biskyrmions.** (a) under-focused and (b) over-focused images of the biskyrmions. (c) The corresponding in-plane magnetization distribution map with a color wheel in the inset. (d) An enlarged single biskyrmion showing the in-plane, vortex-like spin configuration. Under-focused (e) and over-focused (f) Lorentz TEM image of a single biskyrmion. The scale bar is 100 nm. In (d-f), plus (+) and minus (-) indicate the magnetic helicity; that is, the clockwise- and counter-clockwise rotating directions of in-plane magnetizations around the core, respectively.

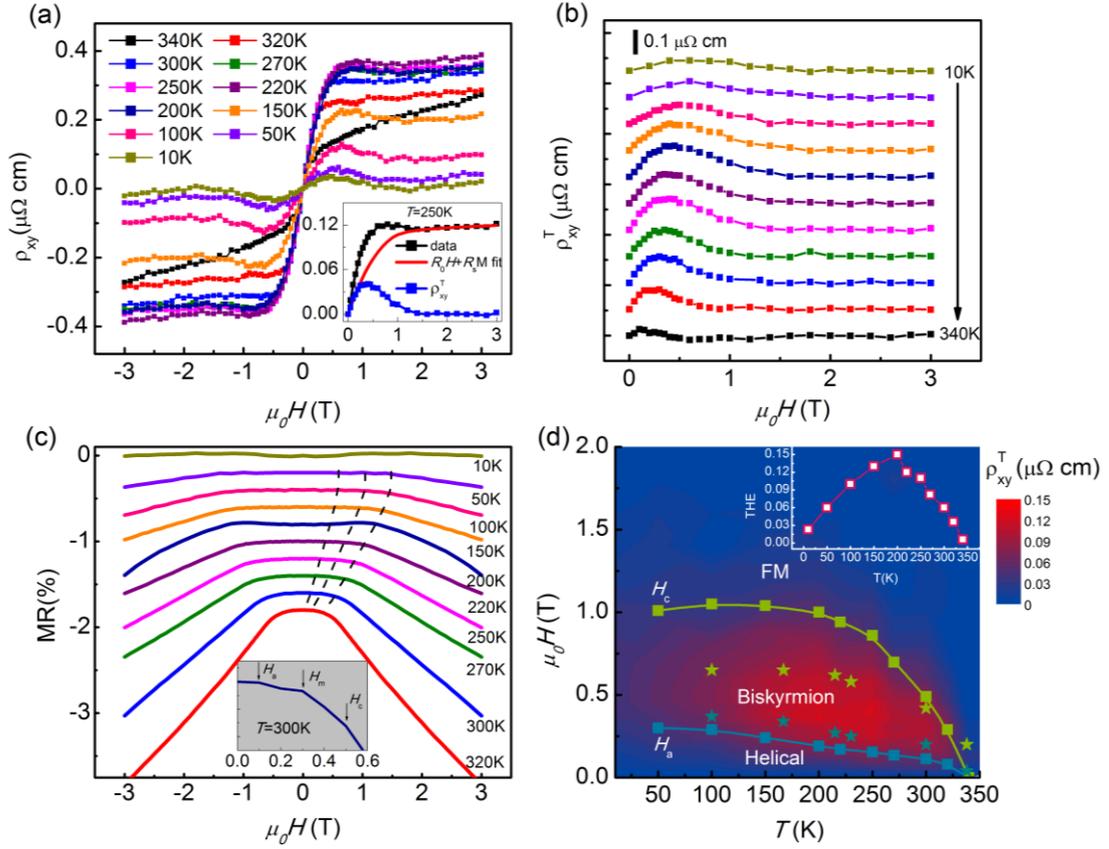

**Figure 4. Topological Hall effect, magnetoresistivity and corresponding magnetic phase diagrams.** (a) Hall resistivity ($\rho_{xy}$) measured at various temperatures from 10 to 340 K. Inset (a): The representative $\rho_{xy}$-$H$ curve (black line) measured at 250 K, calculated $R_0H + R_sM$ curve (red line), and derived $\rho_{xy}^T$ curve (blue line) from the analysis. At high magnetic fields, the calculated result is in accord with the experimental curve. (b) Topological resistivity ($\rho_{xy}^T$) at various temperatures extracted from $\rho_{xy} - T$ curves in (a). (c) Normalized magnetoresistivity measured from 10 to 320 K. The dashed lines indicate the approximate positions of the critical field corresponding to the $H_a$ (blue line), $H_c$ (yellow line) in the $H$-$T$ phase diagram in panel d. (d) The contour mapping of extracted $\rho_{xy}^T$ as a function of the external magnetic field ($H$) and temperature ($T$) for MnNiGa; The blue and yellow lines are the values of $H_a$ and $H_c$, which indicate the phase boundaries between the helical, biskyrmion and field-induced spin collinear FM states. The blue and yellow stars indicate the experimental data based on the *in-situ* Lorentz TEM observations. Inset of (d): maximum values of $\rho_{xy}^T$ as a function of temperature.

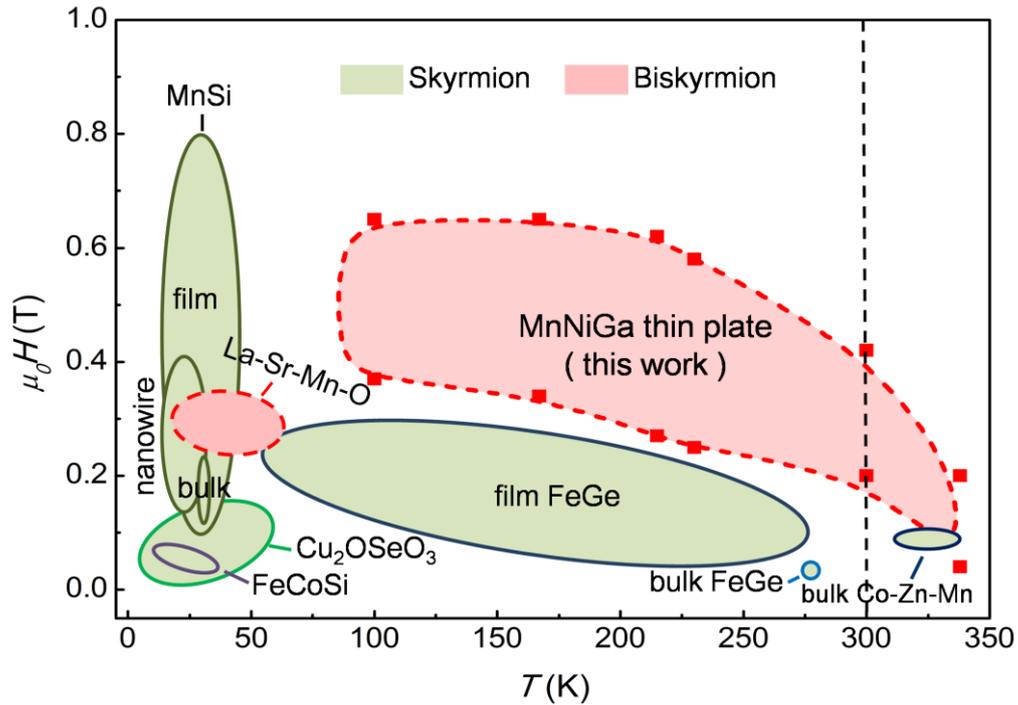

**Figure 5.** Comparison of the region of skyrmion formation in temperature-magnetic field (*T-H*) plane for the magnetic skyrmion- and biskyrmion-hosting materials, respectively. Films and nanowires are also included for reference. The data are taken from the open literatures, including Refs. [1-12],[22],[29-31],[33]. The red solid squares indicate the experimental data based on the *in-situ* Lorentz TEM observations.

**Superstable biskyrmion magnetic naondomains** are expreimental observed for the first time in a hexagonal MnNiGa, a common and easily produced centerosymmetric material. The biskyrmion states in MnNiGa thin plates, as determined by the combination of *in-situ* Lorentz transmission electron microscopy images, magnetoresistivity and topological Hall effect measurements, are surprisingly stable over a broad temperature range of 100 K to 340 K.

*Wenhong Wang\*, Ying Zhang\*, Guizhou Xu, Licong Peng, Bei Ding, Yue Wang, Zhipeng Hou, Xiaoming Zhang, Xiyang Li, Enke Liu, Shouguo Wang, Jianwang Cai, Fangwei Wang, Jianqi Li, Fengxia Hu, Guangheng Wu, Baogen Shen, Xi-xiang Zhang*

A centrosymmetric hexagonal magnet with superstable biskyrmion magnetic nanodomains in a wide temperature range of 100K to 340K

TOC Figure

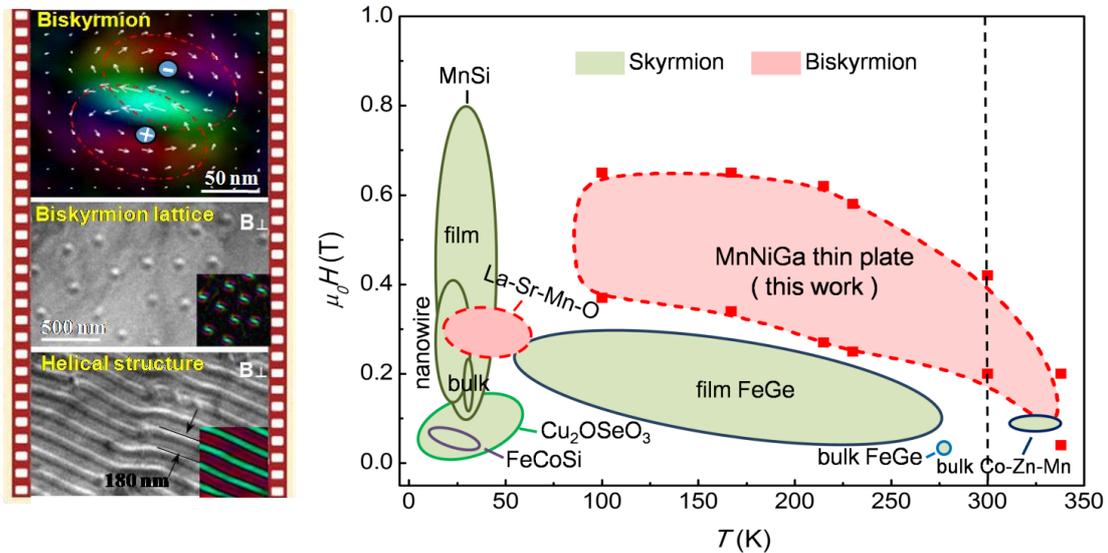